\documentclass[11pt]{article}
\usepackage{amsmath}
\usepackage{amssymb}
\usepackage{verbatim}
\usepackage[pdftex]{graphicx}
\usepackage{palatino}
\usepackage[T1]{fontenc}
\usepackage{microtype}
\usepackage{hyperref}

\begin{document}
\title{Oriented polar molecules in a solid inert-gas matrix: 
a proposed method for measuring the electric
dipole moment of the electron}
\author{A.C. Vutha\footnote{Department of Physics, University of Toronto}
\footnote{Corresponding author: vutha@physics.utoronto.ca}, 
M. Horbatsch\footnote{Department of Physics and Astronomy, York Univerity}, 
E.A. Hessels\footnote{Department of Physics and Astronomy, York Univerity}}

\maketitle

\abstract{
We propose a very sensitive method for measuring
the electric dipole moment of the electron 
using polar molecules embedded in a cryogenic solid matrix of 
inert-gas 
atoms. 
The polar molecules can be oriented in the 
$\hat{\rm{z}}$
direction by an applied electric field,
as has recently been demonstrated by Park, 
et al.
[Angewandte Chemie {\bf 129}, 1066 (2017)].
The trapped molecules are prepared into a 
state which has its
electron spin perpendicular to 
$\hat{\rm{z}}$, 
and a
magnetic field along 
$\hat{\rm{z}}$
causes precession of this spin.
An electron electric dipole moment
$d_e$
would affect this 
precession due to the 
up to 
100~GV/cm
effective electric field 
produced by the 
polar molecule.
The large number of polar molecules that can be embedded in a matrix,
along with the expected long coherence times for the precession, 
allows for the possibility of measuring 
$d_e$
to an accuracy that surpasses current measurements by many orders 
of magnitude. 
Because the matrix can inhibit molecular rotations and lock the orientation
of the polar molecules, 
it may not be necessary to have an electric 
field present during the precession.
The proposed technique can be applied using a 
variety of polar molecules and inert gases,
which, along with other experimental variables,
should allow for careful study of systematic uncertainties
in the measurement.
}

\section{Introduction}

Extensions of the Standard Model of particle physics are 
necessary to explain the presence of dark matter and to 
explain the asymmetry between matter and antimatter in 
the universe. 
The Standard Model predicts that the electric dipole 
moment of the electron, 
$d_e$, 
is along the direction of its spin,
and is probably of order 
10$^{-40}~e$~cm
\cite{PhysRevD.89.056006,
pospelov1991electric,
Booth1993}, 
whereas most extensions 
(for example,
supersymmetric theories
\cite{PhysRevD.90.055006,
PhysRevD.87.113002})
predict a much larger value for 
$d_e$.  

The two most precise measurements of 
$d_e$
are by the ACME collaboration 
\cite{ACME2014}
(who use a beam of metastable 
$^{232}$Th$^{16}$O, 
and measure 
$d_e=-$2.1(4.5)$\times$10$^{-29}~e~$cm), 
and the 
JILA
collaboration
\cite{Cornell2017}
(who use trapped metastable 
$^{180}$Hf$^{19}$F$^+$ 
ions and measure 
$d_e=$ 0.9(7.8)$\times$10$^{-29}$ $e~$cm).
The results of these measurements are 
both consistent with zero and their weighted average
gives 
$d_e= -$1.4(3.9)$\times$10$^{-29}~e$~cm, 
which sets 
a 
90$\%$ 
confidence interval of
$|d_e|<$ 7$\times$10$^{-29}~e$~cm.
A stronger limit
(or a nonzero measurement of $d_e$)
is necessary to guide Standard-Model extensions.

Measurements of 
$d_e$
are performed by watching electron spins 
precess within a magnetic field, 
and measuring any change in this precession rate 
due to the presence of an electric field.
The angle through which they precess is given by
\begin{equation}
\phi = (g \mu_B B \pm d_e E_{\rm eff}) T / \hbar,
\label{eq:phi} 
\end{equation}
where 
$g \mu_B$ 
is the magnetic moment of the electron,
$B$ 
is the applied magnetic field,
$E_{\rm eff}$ 
is the effective electric field that the electron experiences
inside the molecule or molecular ion,
and 
$T$ 
is the time that the electron is precessing.
The 
$\pm$
signs correspond to the cases for which the 
electric and magnetic fields are oriented 
in parallel and antiparallel directions.

The accuracy of $d_e$ from such a measurement, 
assuming the measurement is statistically limited by shot-noise,
is
\begin{equation}
\delta d_e = \frac{\hbar}{2 E_{\rm eff}\sqrt{N} T},
\label{eq:uncert} 
\end{equation} 
where 
$N$ 
is the total 
(integrated) 
number of electrons whose precession
is detected.
The value of 
$E_{\rm eff}$ 
is between 
10 and 100~GV/cm for 
most polar molecules 
(and polar molecular ions) 
used for electron 
electric-dipole-moment
studies,
and therefore greatly improved measurements can only be 
obtained by large improvements in 
$N$ 
or in 
$T$.
The ACME experiment
\cite{ACME2014} 
uses an 
$N$ 
of order 
10$^{10}$, 
and a 
$T$ 
of order 
1~ms.
The 
JILA 
measurement
\cite{Cornell2017,Cornell2011} 
uses an 
$N$ 
of order 
10$^6$,
and a 
$T$ 
of approaching 
1~s.

Improvements in 
$N$
and in 
$T$
should be possible for
both the molecular ion and
the neutral molecule measurements.
For molecular ion experiments, 
the number of trapped ions seem 
to be practically limited to be less than approximately 
10$^5$ 
due to interactions 
between co-trapped ions. 
Laser cooling and trapping experiments of neutral polar molecules 
are planned by several groups 
(e.g.,
with 
YbF
\cite{Tarbutt2013}
and
TlF
\cite{Hunter2012}), 
but, 
even
assuming the same performance from a molecular trap 
as an atom trap, 
the maximum molecule number is unlikely to exceed 
10$^8$ 
molecules per trap cycle.
Beam experiments with a
$T$ 
of approaching 
1~s  
could potentially reach 
$N$ 
of 
10$^{10}$, 
if the considerable experimental difficulties of maintaining 
beam collimation and field control over tens of meters
can be achieved.

We propose here a method which will allow for a much larger 
$N$, 
while maintaining a large value of  
$T$. 
The method involves embedding polar molecules in an
inert-gas 
matrix. 
We refer to this method as 
EDM$^{\bf 3}$ 
(Electric Dipole Measurements using Molecules within a Matrix).
The method exploits the fact that the 
inert-gas
matrices are transparent, 
and that the influence of the matrix on the 
molecule is small enough to still allow for 
state-preparation
and
detection 
techniques similar to those used  
in molecular beam measurements of
$d_e$. 
The number of embedded polar molecules could range
from 
10$^{12}$ to 10$^{16}$,
or more. 
We expect that a 
$T$ 
of approaching 
1~s
should be possible
given that 
such coherence times have been demonstrated in 
bulk materials at 77 kelvin 
\cite{bar2013solid}
(and we expect longer coherence times
with 
inert-gas
matrices which have larger lattice 
separations and can be grown with ultrahigh
purity),
and given that 
second-long coherence times were demonstrated 
for cesium atomic spins in 
solid-helium 
crystals 
\cite{Arndt1995, Kanorsky1996}. 
With a measurement cycle time of one second and 
a 
one-month-long 
measurement,  
$N$ 
would be 
10$^{18}$ 
to 
10$^{22}$,
or more. 
This would lead to a statistical measurement accuracy
of 
$\delta d_e$ = $\sim$10$^{-35}$ to $\sim$10$^{-37} e$ cm,
which represents an improvement of between seven and nine
orders of magnitude when compared to the current
uncertainty on 
$d_e$.

The 
EDM$^{\bf 3}$
method requires the embedded polar 
molecules to be oriented within 
the 
inert-gas
matrix. 
Only recently
\cite{park2017brute}
has it been demonstrated that polar molecules can be
fully oriented in an inert-gas matrix
by the application of electric fields of between 
1
and 
3~MV/cm,
using the novel technique
\cite{shin2013generation}
of 
ice-film 
nanocapacitors. 
Previous to this achievement,
it was thought
\cite{
lemeshko2013manipulation}
that a completely oriented sample would not 
be feasible.

Furthermore,
molecules trapped in a matrix can have 
their rotations inhibited by the forces 
between the 
ions
that make up the molecule 
and the polarizable 
inert-gas
atoms that are their
nearest neighbors in the matrix.
Theoretical understanding of this
inhibition of rotations 
(which can lock molecules into
librator states instead of their 
usual rotational states)
has been studied using 
Devonshire octahedral
model potentials
\cite{Flygare,Kiljunen2005a}.
The model
describes the motion of the trapped molecules 
in terms of hindered rotations and librations,
and shows that the rotational spectrum includes trapped librator states for strong coupling.
Here, 
strong coupling occurs for large values of 
the ratio of the 
potential barrier that hinders
rotation 
(caused by the 
inert-gas--molecule 
interactions) 
divided by the rotational constant 
of the molecule.
The librator states may allow
the polar molecules to remain 
oriented, 
even after the applied electric field is 
turned off.

There are several advantages of measuring
$d_e$ in a matrix, 
as opposed to in a solid
such as gadolinium-iron garnet
\cite{PhysRevLett.95.253004,
PhysRevD.91.102004,
PhysRevA.66.022109}).
The spacing between 
inert-gas 
atoms in the matrix is large,
leading to only small perturbations on
the embedded molecules to be studied.
Since
the matrix is transparent,
standard 
spectroscopy techniques
can be used in studying the 
embedded molecules. 
Also, 
the matrix allows for the large 
$E_{\rm eff}$
available in polar molecules,
and,
in addition,
one can  
easily change the species 
and density of the embedded 
molecules.
With 
EDM$^{\bf 3}$,
$d_e$ can be measured 
using the individual molecules
(using methods similar to 
those used for trapped ions
or molecular beams),
rather than observing the 
bulk properties of the solid.
That is, 
EDM$^{\bf 3}$
gives both the advantage of 
measuring 
$d_e$
with
large sample sizes
(as in 
solid-state
measurements),
and 
the advantage of applying
precise,
shot-noise-limited
spectroscopic techniques
(as used in the most
precise measurements
of 
$d_e$
\cite{ACME2014,
Cornell2017}
).

A previous suggestion 
\cite{kozlov2006proposal}
of using 
molecules in an inert-gas 
matrices for measurements of 
$d_e$ 
involved detection of a 
very small magnetization induced
by the presence of  
$d_e$ 
in the 
$E_{\rm eff}$
of the molecule,
assuming that the molecules are 
oriented by an external field.
Here we suggest a more standard 
approach to measuring 
$d_e$ 
using techniques similar to those
used for 
molecular-beam 
measurements of 
$d_e$ 
(see, 
for example,
Ref~\cite{hudson2011improved,
kara2012measurement}).
Other authors have explored 
\cite{weis1997can}
the possibility of using atoms within
an 
inert-gas
matrix to measure 
$d_e$. 
Using atoms instead of molecules has
the advantage of less complicated,
and more easily calculated, 
quantum states, 
but has the drawback of the much smaller
$E_{\rm eff}$ 
(of approximately
100 
times the applied field
\cite{nataraj2008intrinsic}).
Nonetheless, 
the possibility of using atoms may 
have to be reevaluated in light of the 
much larger applied electric fields 
recently demonstrated in 
Ref.~\cite{park2017brute},
although 
these fields will still lead to
an
$E_{\rm eff}$
of approximately 
100~MV/cm,
compared to the 
approximately
100~GV/cm fields
obtained with polar molecules.

Some details of the proposed 
EDM$^{\bf 3}$
method are 
outlined in the following section.
However, 
many of the steps will need further development
and experimental verification to prove the viability and 
strength of the  
EDM$^{\bf 3}$ 
method for 
determining  
$d_e$.
The methods described are also well suited to
nuclear electric dipole measurements, 
as nuclear spin coherence times of well over
1~s 
can be obtained at low temperatures
in bulk material
\cite{zhong2015}, 
and we would expect even longer times
within  
inert-gas 
matrices.

\section{Components for the EDM$^{\bf3}$ method}

\subsection{Inert gas solids}

The solid state of inert gases has been studied for 
over a century, 
and their properties 
(see,
for example,
\cite{Pollack1964})
are well known. 
The inert solids most commonly used 
for matrix studies 
(Ne, 
Ar, 
Kr 
and Xe)
all have face-centered-cubic 
(fcc)
structure,
with the cube sizes being
4.5, 
5.3,
5.6,
and 
6.1~Angstroms,
respectively.
They have melting points of 
24.5,
83.8,
115.9,
and 
165.1~K,
and all have extremely low vapor pressures 
if cooled to temperatures of a few kelvin or lower.
Single crystals of a cubic centimeter or 
larger
\cite{Endoh1975, Berne1966}
have been produced. 
Bulk solids have been shown to sustain fields of over 
10~kV/cm
\cite{Eshchenko2002}
without arcing, 
and fields as large as 
3~MV/cm 
have been successfully applied to thin films
\cite{park2017brute}.

\subsection{The study of molecules trapped in an inert-gas solid}

Inert-gas matrices have proven to be very useful 
(for example,
Refs.~\cite{Mann1966,
VanZee1977,
Dubost1976,
Knight1980,
Mason1971})
for studying molecules trapped within them 
-- particularly reactive molecules, 
since the individual molecules can be isolated
by the matrix. 
Because the lattice is transparent,
the trapped molecules can be excited with
lasers and fluorescence can be observed.
Although the presence of the 
inert-gas matrix broadens the spectral lines,
resolutions of 
0.1
to
1~cm$^{-1}$,
or better,
(for example, 
Refs.~\cite{Dubost1976}, 
\cite{lang1991matrix},
and
\cite{park2017brute})
are still possible.
The matrix can suppress
\cite{Kiljunen2007,Kiljunen2006a} 
molecular rotations, 
leaving molecules librating about a preferred direction.
Electron spin resonance spectroscopy
(see, 
for example,
\cite{Knight1983})
has been routinely used to study
matrix-trapped 
molecules.
Optically pumped rubidium atoms within an inert-gas
matrix
\cite{Kanagin2013}
show that electron spin polarization can be maintained for 
a coherence time of up to 
0.1~s.

\subsection{Oriented polar molecules within an 
inert-gas 
lattice}

Polar molecules are particularly suited to being 
trapped in an inert-gas matrix.
The strong interaction between the polarizable 
inert-gas
atoms of the matrix and the two charges within the 
(typically ionically bonded)
polar molecule
serves to clamp the position of the polar molecule, 
as well as inhibiting its rotational motion. 

Recently,
Park, 
et al., 
\cite{park2017brute}
have demonstrated that embedded polar molecules within
solid 
Ar,
can be fully oriented by an
externally applied electric field.
They demonstrated complete orientation with
CH$_2$O, 
HCl, 
and 
H$_2$O
molecules 
embedded at a concentration of approximately 
1:1000 
within a
7-kelvin
Ar solid.
The field required for complete orientation 
was approximately
1~MV/cm,
and they showed that the 
fractional orientation of the sample follows a thermodynamic 
model which predicts that the applied electric field
required is proportional to the temperature 
of the matrix divided by the dipole moment
of the polar molecule. 
At colder temperatures,
such as those available from 
a pumped liquid helium 
($\sim$1~K)
or a dilution refrigerator
($\sim$100~mK),
the field needed for complete 
alignment would be expected to be 
correspondingly smaller.

Park, 
et al., 
\cite{park2017brute} 
produced the  
1~MV/cm 
fields 
using the recently developed technique
\cite{shin2013generation}
of 
ice-film 
nanocapacitors,
in which
cesium ions are deposited
on top of a 
300-nm-thick 
film 
(grown on a cryogenic platinum
substrate) 
that consisted of solid argon 
(with embedded polar molecules) 
sandwiched between
two layers of 
D$_2$O.
The number of polar molecules in their 
sample was approximately
10$^{13}$ per mm$^2$, 
and,
at this concentration,
they found that monomers strongly dominate
over multimers.
With areas of a square centimeter, 
or more, and with scaling to thicker
films, 
one could imagine doing an 
EDM$^{\bf 3}$
experiment with 
$\ge$10$^{16}$ 
polar molecules.

\subsection{Choice of polar molecule}

To measure 
$d_e$,
one needs a molecule which has 
at least one unpaired electron spin
(which will precess under the 
combined magnetic and electric fields).
For large
$E_{\rm eff}$,
one needs a 
polar molecule  
in which the unpaired electron is as close
as possible to one of the nuclei.
This proximity to the nucleus
requires a heavy atom 
with an unpaired
electron in an
s-orbital.
Molecules that satisfy these criteria 
typically tend to be chemically reactive. 
Therefore, 
they are usually produced as molecular beams 
in a vacuum by methods ranging from 
laser ablation of solid precursors 
(ablated into an inert or reactive carrier gas stream
\cite{tarbutt2002jet,
hutzler2011cryogenic}) 
to chemical reactions at high temperatures 
\cite{west2017improved}. 
The 
EDM$^{\bf 3}$ 
method is compatible with the production 
of molecules using such methods, 
as these molecular beams can be mixed 
into an 
inert-gas 
stream, 
which then impinges onto a 
cryogenic surface to grow an
inert-gas crystal with embedded molecules. 

An important advantage of the 
EDM$^{\bf 3}$ scheme is that the 
trapping method is generic, 
and can be used with a wide variety of 
polar molecules, 
irrespective of their chemical stability. 
In particular, 
a range of molecules with similar chemical properties 
but large differences in their values of 
$E_{\rm eff}$ 
(e.g., 
YbF
and
HgF, 
as shown in 
Table~\ref{tbl:molecules}), 
can be used as a powerful means of distinguishing 
systematic effects from a real 
$d_e$
signal.

\begin{table}[t]
\centering
\caption{
\label{tbl:molecules} 
$E_{\rm eff}$ 
for some candidate molecules that 
could be used with the 
EDM$^{\bf 3}$
method.
Other molecules,
including chlorides and oxides,
would also be suitable for 
EDM$^{\bf 3}$.
}
\begin{tabular}{ccc}
\\
\hline
\hline
molecule&$E_{\rm eff}$ (GV/cm)&Ref.\\
\hline
YbF&23&\cite{PhysRevA.90.022501}  \\
HgF&115&\cite{PhysRevLett.114.183001} \\
HgCl&114&\cite{PhysRevLett.114.183001} \\
HgBr&109&\cite{PhysRevLett.114.183001} \\
HgI&109&\cite{PhysRevLett.114.183001}\\
RaF&52&\cite{PhysRevA.93.062506} \\
WC&54&\cite{Lee2005} \\
\hline
\hline
\end{tabular}
\end{table}

A wish list for a molecule to be used in the 
EDM$^{\bf 3}$ 
scheme includes a large value of 
$E_{\rm eff}$ 
and a large electric dipole moment. 
The latter is important to facilitate
orientation in a smaller applied electric field.
Additionally, 
a large difference between the 
ground-state electric dipole moment of the 
molecule and that in an excited state 
will lead to large  
Stark shifts of optical transitions
between the two states, 
which could be used for selectively
detecting molecules along a particular orientation.
A large moment of inertia for the molecule is 
also desirable, 
since it will help to lock the molecule into 
a librator state.
Table~\ref{tbl:molecules}
lists 
$E_{\rm eff}$ 
for some candidate molecules
that might be used in an 
EDM$^{\bf 3}$ 
experiment.
The list includes the radioactive molecule
RaF 
since the 
EDM$^{\bf 3}$ 
method is well-suited for to experiments with 
radioactive molecules, and is compatible wiht 
their production from 
rare-isotope
beam sources.

\subsection{The time sequence for EDM$^{\bf 3}$}

The envisioned 
EDM$^{\bf 3}$
experiment would use cold polar molecules 
from a cryogenic 
buffer-gas 
beam source sprayed onto a 
cryogenic surface along with a steady stream of 
an inert gas
(likely, neon or argon). 
Once polar molecules are oriented within the 
inert-gas 
matrix,
the same set of molecules can be repeatedly used for 
EDM$^{\bf 3}$ 
measurements.

\begin{figure}
\centering
\includegraphics[width=3in]{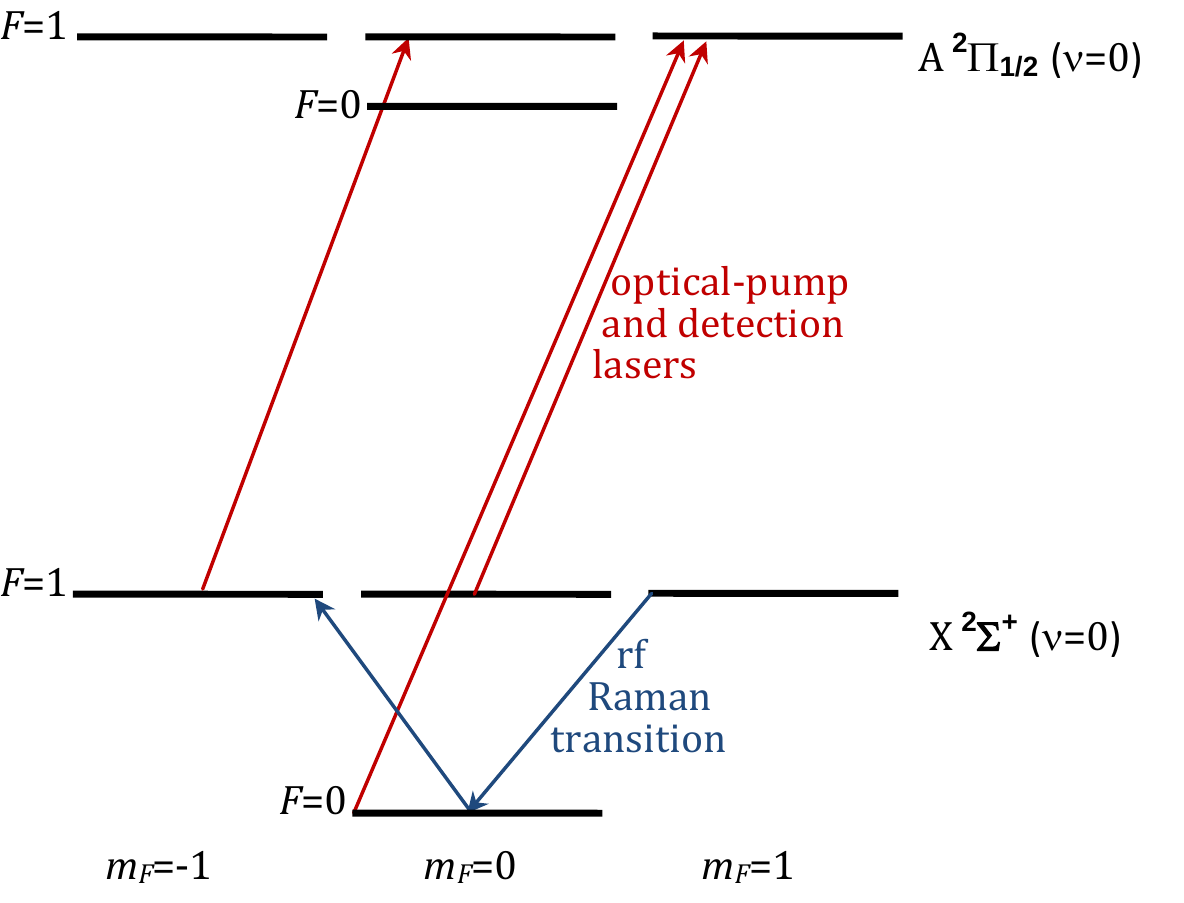}
\caption{\label{fig:YbFlevels} (Color online) 
Energy-level diagram for relevant levels of YbF, 
as an illustration of how  
EDM$^{\bf 3}$ could be implemented.
The diagram is not to scale.
} 
\end{figure}

For simplicity, 
we describe a straightforward measurement sequence 
that would allow for an
EDM$^{\bf 3}$
measurement of 
$d_e$, 
and we use the 
YbF 
molecule,
to illustrate the sequence.
The relevant energy structure for the oriented
YbF
molecule
is shown in 
Fig.~\ref{fig:YbFlevels}.

At the cryogenic temperatures of the matrix, 
the molecules will be  
thermalized into the ground
vibrational state, 
as well as into the 
lowest-energy 
librational state.
(For the case of strongly-oriented molecules, 
the usual rotational states are 
replaced by more widely spaced
librational states.)
The 
X~$^2\Sigma^+$
ground state of 
$^{174}$YbF,
has a total nuclear spin
of 
1/2 
and a total electron spin of 
1/2,
giving hyperfine states with 
$F$=0
and
$F$=1,
as shown in 
Fig.~\ref{fig:YbFlevels}.
In the first step of the measurement sequence, 
the oriented 
molecules are optically pumped 
into the 
$F$=1, $m_F$=$1$ 
state
using circularly polarized light 
tuned to the 
X~$^2\Sigma^+$($\nu$=0)-to-A~$^2\Pi_{1/2}$($\nu$=0)
transition. 
A pulsed laser 
(or other broadband source)
is used to broaden the spectrum of the light
so that it can communicate to
both hyperfine levels, 
and to allow for any inhomogeneous 
broadening that might be caused by the matrix.
Radiative decay paths to excited vibrational states are
expected to be small, 
and molecules in the excited vibrational states
will decay back to the ground vibrational state on the 
10-ms time scale 
(or faster, 
due to the influence of the matrix).

The next step in the measurement sequence is a
$\pi$/2 pulse of 
rf 
driving the 
$F$=1, $m_F$=$1$ 
to
$F$=1, $m_F$=$-1$ 
Raman
transition 
(see
Fig.~\ref{fig:YbFlevels}).
This pulse creates an equal 
coherent superposition of the 
$F$=1, $m_F$=$1$
and 
$F$=1, $m_F$=$-1$
states, 
which has an electron spin angular momentum 
oriented in the 
$\hat{x}$
direction.
This spin will precess for a time
$T$
due to an applied magnetic field in the 
$\hat{z}$
direction
and 
(if 
$d_e$ 
is nonzero)
due to the 
$E_{\rm eff}$
aligned with 
$\hat{z}$,
as described by
Eq.~(\ref{eq:phi}).
Spin precession is observed 
routinely in 
matrix-trapped 
molecules
in electron spin resonance spectroscopy
(see, 
for example,
\cite{Knight1983}).  
The magnitude of the magnetic field
used for the precession will depend
on a number of factors, 
including the inhomogeneity of the
field
(caused, 
for example,
by stray fields due to 
the substrate, 
or due to imperfect
shielding)
and the possible variation
in the 
$g$ 
factor 
(for example,
at different sites in the 
lattice or due to lattice imperfections).
The precession time
$T$ 
could be approximately 
1~s.
The best 
$T$
to use would depend on the 
measured coherence time for the electron spin,
but 
$\sim$1~s 
seems possible given that 
second-long coherence times were reported 
for cesium atomic spins in 
solid-helium 
crystals 
\cite{Arndt1995, Kanorsky1996}, 
and that 
0.6~s 
coherence time were seen 
at 77 kelvin for
nitrogen-vacancy color centers in diamond
\cite{bar2013solid}.

After precession,
the spin direction could be detected by 
a combination of another 
rf 
Raman 
pulse, 
followed by a 
circularly-polarized 
detection laser
and observation of the resulting fluorescence.
A
3$\pi$/2
rf
Raman
pulse would transfer 
an electron spin oriented along 
$\hat{x}$
back into the 
$F$=1, $m_F$=$1$
state,
whereas it would transfer 
an electron spin oriented along 
$\hat{y}$
into the 
$F$=1, $m_F$=$-1$
state.
Right-circular-polarized 
($\sigma^+$)
laser light
(see
Fig.~\ref{fig:YbFlevels})
only excites molecules in the 
$m_F$=$-1$
state
(i.e.,
the 
$m_F$=$1$
state is a dark state), 
and,
therefore,
observing the 
resulting fluorescence 
gives a measurement of the 
precession angle.
An alternative to 
rf
Raman
pulses is to use optical pumping with 
laser pulses that are synchronized with
the spin precession 
\cite{PhysRevA.52.3521}.

To determine 
$d_e$,
the measurement needs to be repeated with
the relative direction of the electric and 
magnetic field reversed, 
to see if there is a difference in the precession.
The details of the measurement technique employed in
an 
EDM$^{\bf 3}$ 
experiment would, 
of course,
depend on the molecule used,
and,
more importantly, 
on the techniques needed to control 
possible systematic effects. 

\subsection{Orientation without an electric field}

Because the interaction between the matrix and the polar 
molecule can inhibit rotation
\cite{Flygare,Kiljunen2005a,Kiljunen2007,Kiljunen2006a}, 
it may be possible to find a polar molecule 
(and an inert gas)
for which the orientation is sufficiently well locked
to allow the orientation of the polar molecules to 
remain,
after the external electric field that oriented them
is turned off.
The locked orientation might allow for an 
EDM$^{\bf 3}$ 
measurement in the absence of an applied electric field,
which would minimize any systematic effects
associated with this electric field.

Furthermore, 
it may not even be necessary to apply an electric
field to initially orient the molecules. 
In a single 
inert-gas 
crystal, 
the polar molecules prefer to align along
a number of axes relative the the fcc matrix
(as determined by the interaction between the dipole 
and the neighboring 
inert-gas atoms in the crystal).
If the molecular orientation is sufficiently locked,
molecules that start in one orientation will remain 
in that orientation,
at least for the one second required 
to complete a measurement of 
$d_e$. 
One could selectively perform a measurement on 
the molecules that are oriented along the 
$+\hat{z}$
direction
(chosen to be one of the preferred axes) 
by selectively detecting only molecules in this orientation, 
using the Stark shifts from an electric field 
applied during the detection step. 
The applied electric field 
need only be present during optical excitation of
this detection
step 
(and does not need to be present during 
the precession of the electron spin).
Selective detection of 
$+\hat{z}$-oriented
molecules 
result from the fact that the 
Stark-shift rates are different for
different orientations 
of the molecule.
(For an applied field of 
10~kV/cm,
a typical polar molecule 
experiences shifts of between
approximately 
$-$10~GHz  
and
$+$10~GHz,
depending on the orientation of the
molecule relative to the applied field.)
The detection laser would 
be tuned to 
the 
Stark-shifted 
frequency of the 
$+\hat{z}$-oriented
molecules to select only these molecules,
and therefore only record the precession for 
those electrons that experience an
$E_{\rm eff}$
in this direction.

\section{Potential advantages of the EDM$^{\bf 3}$ method}

The main advantage of the 
EDM$^{\bf 3}$ 
method is the large number of molecules
that can be embedded into the matrix,
and the long interaction times possible for 
these trapped molecules, 
along with the large effective electric field 
that is available with polar molecules.
That is, 
EDM$^{\bf 3}$
can allow for large 
$N$, 
$T$
and 
$E_{\rm eff}$
in 
Eq.~(\ref{eq:uncert}),
allowing for a large improvement in
the statistical uncertainty
of a 
$d_e$
measurement.

Other potential advantages of 
EDM$^{\bf 3}$
involve the control
of systematic effects.
Because the crystal is built out of an inert gas,
it can be produced with ultrahigh purity compared to 
ordinary
solid-state
hosts.
Additionally,
the measurement can easily be repeated
with different inert gases,
or different polar molecules, 
and with different distances between 
the polar molecules,
in order to study systematic effects.
Similarly, 
the substrate
on which the gas is frozen,
which could be an important contributor to 
decoherence and systematic effects,
could be varied.
The substrate 
(and any impurities in the substrate) 
will likely have to be within  
micrometers 
of the polar molecules. 
Magnetic Johnson noise from conducting substrates 
could be important 
(for this, low temperatures and resistive substrates 
would be better), 
and impurities 
(including nuclear spins) 
in the substrate can lead to decoherence. 
A material such as isotopically pure silicon
or diamond 
might be necessary in order to obtain a 
mechanically and magnetically 
clean substrate.

Another advantage of
EDM$^{\bf 3}$
is the small volume
in which the molecules are contained.
The small volume will allow
for efficient collection and detection of fluorescence. 
It will also allow for easier magnetic shielding,
and could allow for the magnetic coils producing the
fields to be mechanically rotated 
to reverse the field 
(in addition to reversing the field by reversing 
the current).
The thin film envisioned for the measurement
would minimize the effect of magnetic fields caused
by leakage currents.
These fields, 
if they have a component along 
$\hat{z}$,
can mimic the effect of an electron electric dipole
moment.
Here, 
this effect should be minimal, 
since only a small amount of charge 
is needed to create the fields,
and, 
even if it did leak across the sample, 
it would be expected to take an approximately
direct path across the film.
Thus, 
any potential leakage current would be 
small, 
and the migrating charges would be expected to travel
almost parallel to 
the 
$z$
direction, 
leading to an insignificant 
component of magnetic field along
$\hat{z}$.

The possibility of having the precession 
take place without an applied
electric field could be a powerful weapon against 
electric-field induced systematic effects, 
as well as providing orientation reversal 
without electric field reversal 
(even for molecules that do not have 
$\Omega$-doublets
\cite{demille2001search,PhysRevA.87.052130}), 
simply by mechanical rotation of the oriented 
frozen gas sample. 

\section{Conclusion}

We propose the use of oriented polar molecules in 
inert-gas 
matrices as a pathway to improving 
the experimental limit on the 
electric dipole moment of the electron. 
Our proposed method relies on demonstrated techniques, 
and holds tremendous promise for improving the statistical 
sensitivity of an electron 
electric dipole moment experiment. 
At the same time, 
the method offers a number of advantageous 
features for suppressing systematic errors.
Improved measurements of this moment
would provide strong input for 
beyond-the-Standard-Model
theories of high-energy physics.

\bibliography{edmCubed}

\begin{thebibliography}{10}

\bibitem{PhysRevD.89.056006}
M.~Pospelov and A.~Ritz.
\newblock {CKM benchmarks for electron electric dipole moment experiments}.
\newblock {\em Phys. Rev. D}, 89:056006, 2014.

\bibitem{pospelov1991electric}
M.~E. Pospelov and I.~B. Khriplovich.
\newblock {Electric dipole moment of the W boson and the electron in the
  Kobayashi-Maskawa model}.
\newblock {\em Yadernaya Fizika}, 53:1030, 1991.

\bibitem{Booth1993}
M.~Booth.
\newblock The electric dipole moment of the $\rm{W}$ and electron in the
  standard model.
\newblock {\em arXiv hep-ph/9301293}, 1993.

\bibitem{PhysRevD.90.055006}
T.~Ibrahim, A.~Itani, and P.~Nath.
\newblock {Electron electric dipole moment as a sensitive probe of PeV scale
  physics}.
\newblock {\em Phys. Rev. D}, 90:055006, 2014.

\bibitem{PhysRevD.87.113002}
D.~McKeen, M.~Pospelov, and A.~Ritz.
\newblock {Electric dipole moment signatures of PeV-scale superpartners}.
\newblock {\em Phys. Rev. D}, 87:113002, 2013.

\bibitem{ACME2014}
J.~Baron, W.~C. Campbell, D.~DeMille, J.~M. Doyle, G.~Gabrielse, Y.~V.
  Gurevich, P.~W. Hess, N.~R. Hutzler, E.~Kirilov, I.~Kozyryev, B.~R.
  O{\lq}Leary, C.~D. Panda, M.~F. Parsons, E.~S. Petrik, B.~Spaun, A.~C. Vutha,
  and A.~D. West.
\newblock Order of magnitude smaller limit on the electric dipole moment of the
  electron.
\newblock {\em Science}, 343:269, 2014.

\bibitem{Cornell2017}
W.~B. Cairncross, D.~N. Gresh, M.~Grau, K.~C. Cossel, T.~S. Roussy, Y.~Ni,
  Y.~Zhou, J.~Ye, and E.~A. Cornell.
\newblock A precision measurement of the electron{\lq}s electric dipole moment
  using trapped molecular ions.
\newblock {\em Phys. Rev. Lett.}, 119:153001, 2017.

\bibitem{Cornell2011}
A.~E. Leanhardt, J.~L. Bohn, H.~Loh, P.~Maletinsky, E.~R. Meyer, L.~C.
  Sinclair, R.~P. Stutz, and E.~A. Cornell.
\newblock High-resolution spectroscopy on trapped molecular ions in rotating
  electric fields: A new approach for measuring the electron electric dipole
  moment.
\newblock {\em Journal of Molecular Spectroscopy}, 270:1, 2011.

\bibitem{Tarbutt2013}
M.~R. Tarbutt, B.~E. Sauer, J.~J. Hudson, and E.~A. Hinds.
\newblock {Design for a fountain of YbF molecules to measure the electron's
  electric dipole moment}.
\newblock {\em New Journal of Physics}, 15:053034, 2013.

\bibitem{Hunter2012}
L.~R. Hunter, S.~K. Peck, A.~S. Greenspon, S.~S. Alam, and D.~DeMille.
\newblock {Prospects for laser cooling TlF}.
\newblock {\em Physical Review A}, 85:012511, 2012.

\bibitem{bar2013solid}
N.~Bar-Gill, L.~M. Pham, A.~Jarmola, D.~Budker, and R.~L. Walsworth.
\newblock Solid-state electronic spin coherence time approaching one second.
\newblock {\em Nature communications}, 4:1743, 2013.

\bibitem{Arndt1995}
M.~Arndt, S.~I. Kanorsky, A.~Weis, and T.~W. H\"{a}nsch.
\newblock {Long electronic spin relaxation times of Cs atoms in solid $^4$He}.
\newblock {\em Phys. Rev. Lett.}, 74:1359, 1995.

\bibitem{Kanorsky1996}
S.~I. Kanorsky, S.~Lang, S.~L\"{u}cke, S.~B. Ross, T.~W. H\"{a}nsch, and
  A.~Weis.
\newblock {Millihertz magnetic resonance spectroscopy of Cs atoms in
  body-centered-cubic $^4$He.}
\newblock {\em Phys. Rev. A}, 54:R1010, 1996.

\bibitem{park2017brute}
Y.~Park, H.~Kang, and H.~Kang.
\newblock Brute force orientation of matrix-isolated molecules: Reversible
  reorientation of formaldehyde in an argon matrix toward perfect alignment.
\newblock {\em Angewandte Chemie}, 129:1066, 2017.

\bibitem{shin2013generation}
S.~Shin, Y.~Kim, E.~Moon, D.~H. Lee, H.~Kang, and H.~Kang.
\newblock Generation of strong electric fields in an ice film capacitor.
\newblock {\em The Journal of Chemical Physics}, 139:074201, 2013.

\bibitem{lemeshko2013manipulation}
M.~Lemeshko, R.~V. Krems, J.~M. Doyle, and S.~Kais.
\newblock Manipulation of molecules with electromagnetic fields.
\newblock {\em Molecular Physics}, 111:1648, 2013.

\bibitem{Flygare}
W.~H. Flygare.
\newblock {Molecular rotation in the solid state. Theory of rotation of trapped
  molecules in rare gas lattices}.
\newblock {\em The Journal of Chemical Physics}, 39:2263, 1963.

\bibitem{Kiljunen2005a}
T.~Kiljunen, B.~Schmidt, and N.~Schwentner.
\newblock {Aligning and orienting molecules trapped in octahedral crystal
  fields}.
\newblock {\em Phys. Rev. A}, 72:053415, 2005.

\bibitem{PhysRevLett.95.253004}
B.~J. Heidenreich, O.~T. Elliott, N.~D. Charney, K.~A. Virgien, A.~W. Bridges,
  M.~A. McKeon, S.~K. Peck, D.~Krause, J.~E. Gordon, L.~R. Hunter, and S.~K.
  Lamoreaux.
\newblock Limit on the electron electric dipole moment in gadolinium-iron
  garnet.
\newblock {\em Phys. Rev. Lett.}, 95:253004, 2005.

\bibitem{PhysRevD.91.102004}
Y.~J. Kim, C.-Y. Liu, S.~K. Lamoreaux, G.~Visser, B.~Kunkler, A.~N. Matlashov,
  J.~C. Long, and T.~G. Reddy.
\newblock New experimental limit on the electric dipole moment of the electron
  in a paramagnetic insulator.
\newblock {\em Phys. Rev. D}, 91:102004, 2015.

\bibitem{PhysRevA.66.022109}
S.~K. Lamoreaux.
\newblock Solid-state systems for the electron electric dipole moment and other
  fundamental measurements.
\newblock {\em Phys. Rev. A}, 66:022109, 2002.

\bibitem{kozlov2006proposal}
M.~G. Kozlov and A.~Derevianko.
\newblock Proposal for a sensitive search for the electric dipole moment of the
  electron with matrix-isolated radicals.
\newblock {\em Phys. Rev. Lett.}, 97:063001, 2006.

\bibitem{hudson2011improved}
J.~J. Hudson, D.~M. Kara, I.~J. Smallman, B.~E. Sauer, M.~R. Tarbutt, and E.~A.
  Hinds.
\newblock Improved measurement of the shape of the electron.
\newblock {\em Nature}, 473:493, 2011.

\bibitem{kara2012measurement}
D.~M. Kara, I.~J. Smallman, J.~J. Hudson, B.~E. Sauer, M.~R. Tarbutt, and E.~A.
  Hinds.
\newblock {Measurement of the electron's electric dipole moment using YbF
  molecules: methods and data analysis}.
\newblock {\em New Journal of Physics}, 14:103051, 2012.

\bibitem{weis1997can}
A.~Weis, S.~Kanorsky, S.~Lang, and T.~W. H{\"a}nsch.
\newblock Can atoms trapped in solid helium be used to search for physics
  beyond the standard model?
\newblock In {\em Atomic Physics Methods in Modern Research}, pages 57--75.
  Springer, 1997.

\bibitem{nataraj2008intrinsic}
H.~S. Nataraj, B.~K. Sahoo, B.~P. Das, and D.~Mukherjee.
\newblock Intrinsic electric dipole moments of paramagnetic atoms: rubidium and
  cesium.
\newblock {\em Physical Review Letters}, 101:033002, 2008.

\bibitem{zhong2015}
M.~Zhong, M.~P. Hedges, R.~L. Ahlefeldt, J.~G. Bartholomew, S.~E. Beavan, S.~M.
  Wittig, J.~J. Longdell, and M.~J. Sellars.
\newblock Optically addressable nuclear spins in a solid with a six-hour
  coherence time.
\newblock {\em Nature}, 517(7533):177, 2015.

\bibitem{Pollack1964}
G.~L. Pollack.
\newblock The solid state of rare gases.
\newblock {\em Reviews of Modern Physics}, 36:748, 1964.

\bibitem{Endoh1975}
Y.~Endoh, G.~Shirane, and J.~Skalyo.
\newblock {Lattice dynamics of solid neon at 6.5 and 23.7 K}.
\newblock {\em Phys. Rev. B}, 11:1681, 1975.

\bibitem{Berne1966}
A.~Berne, G.~Boato, and M.~De~Paz.
\newblock Experiments on solid argon.
\newblock {\em Il Nuovo Cimento B}, 46:182, 1966.

\bibitem{Eshchenko2002}
D.~G. Eshchenko, V.~G. Storchak, J.~H. Brewer, G.~D. Morris, S.~P. Cottrell,
  and S.~F.~J. Cox.
\newblock Excess electron transport and delayed muonium formation in condensed
  rare gases.
\newblock {\em Phys. Rev. B}, 66:035105, 2002.

\bibitem{Mann1966}
D.~E. Mann.
\newblock {Infrared spectra of HCl, DCl, HBr, and DBr in solid rare-gas
  matrices}.
\newblock {\em J. Chem. Phys.}, 44:3453, 1966.

\bibitem{VanZee1977}
R.~J. {Van Zee}, M.~L. Seely, and W.~Weltner.
\newblock {YbH and YbD molecules: ESR and optical spectroscopy in argon
  matrices at 4 K}.
\newblock {\em J. Chem. Phys.}, 67:861, 1977.

\bibitem{Dubost1976}
H.~Dubost.
\newblock {Infrared absorption spectra of carbon monoxide in rare gas
  matrices}.
\newblock {\em Chem. Phys.}, 12:139, 1976.

\bibitem{Knight1980}
L.~B. {Knight Jr} and M.~B. Wise.
\newblock {The generation and ESR investigation of the BeF radical in rare gas
  matrices}.
\newblock {\em J. Chem. Phys.}, 73:4198, 1980.

\bibitem{Mason1971}
M.~G. Mason.
\newblock {Mid- and far-infrared spectra of HF and DF in rare-gas matrices}.
\newblock {\em J. Chem. Phys.}, 54:3491, 1971.

\bibitem{lang1991matrix}
V.~I. Lang and J.~S. Winn.
\newblock {Matrix-isolated OCS: The high resolution infrared spectra of a
  cryogenically solvated linear molecule}.
\newblock {\em The Journal of Chemical Physics}, 94:5270, 1991.

\bibitem{Kiljunen2007}
T.~Kiljunen and B.~Schmidt.
\newblock Alignment and orientation of molecules in matrices.
\newblock {\em Analysis and Control of Ultrafast Photoinduced Reactions},
  87:337, 2007.

\bibitem{Kiljunen2006a}
T.~Kiljunen, B.~Schmidt, and N.~Schwentner.
\newblock {Time-dependent alignment of molecules trapped in octahedral crystal
  fields}.
\newblock {\em J. Chem. Phys.}, 124:164502, 2006.

\bibitem{Knight1983}
L.~B. Knight.
\newblock {ESR investigations of H$_2$O$^+$, HDO$^+$, D$_2$O$^+$, and
  H$_2$$^{17}$O$^+$ isolated in neon matrices at 4 K}.
\newblock {\em J. Chem. Phys.}, 78:5940, 1983.

\bibitem{Kanagin2013}
A.~N. Kanagin, S.~K. Regmi, P.~Pathak, and J.~D. Weinstein.
\newblock {Optical pumping of rubidium atoms frozen in solid argon}.
\newblock {\em Phys. Rev. A}, 88:063404, 2013.

\bibitem{tarbutt2002jet}
M.~R. Tarbutt, J.~J. Hudson, B.~E. Sauer, E.~A. Hinds, V.~A. Ryzhov, V.~L.
  Ryabov, and V.~F. Ezhov.
\newblock {A jet beam source of cold YbF radicals}.
\newblock {\em Journal of Physics B: Atomic, Molecular and Optical Physics},
  35:5013, 2002.

\bibitem{hutzler2011cryogenic}
N.~R. Hutzler, M.~F. Parsons, Y.~V. Gurevich, P.~W. Hess, E.~Petrik, B.~Spaun,
  A.~C. Vutha, D.~DeMille, G.~Gabrielse, and J.~M. Doyle.
\newblock A cryogenic beam of refractory, chemically reactive molecules with
  expansion cooling.
\newblock {\em Physical chemistry chemical physics}, 13:18976, 2011.

\bibitem{west2017improved}
E.~West, J.~Baron, N.~Hutzler, D.~Ang, J.~Haefner, Z.~Lasner, C.~Panda,
  A.~West, D.~DeMille, G.~Gabrielse, and Doyle J.
\newblock {Improved thermochemical beam source of ThO for measuring the
  electric dipole moment of the electron}.
\newblock {\em Bulletin of the American Physical Society}, 62(8):M7.00003,
  2017.

\bibitem{PhysRevA.90.022501}
M.~Abe, G.~Gopakumar, M.~Hada, B.~P. Das, H.~Tatewaki, and D.~Mukherjee.
\newblock {Application of relativistic coupled-cluster theory to the effective
  electric field in YbF}.
\newblock {\em Phys. Rev. A}, 90:022501, 2014.

\bibitem{PhysRevLett.114.183001}
V.~S. Prasannaa, A.~C. Vutha, M.~Abe, and B.~P. Das.
\newblock Mercury monohalides: Suitability for electron electric dipole moment
  searches.
\newblock {\em Phys. Rev. Lett.}, 114:183001, 2015.

\bibitem{PhysRevA.93.062506}
S.~Sasmal, H.~Pathak, M.~K. Nayak, N.~Vaval, and S.~Pal.
\newblock {Relativistic coupled-cluster study of RaF as a candidate for the
  parity- and time-reversal-violating interaction}.
\newblock {\em Phys. Rev. A}, 93:062506, 2016.

\bibitem{Lee2005}
J.~Lee, E.R. Meyer, R.~Paudel, J.L. Bohn, and A.E. Leanhardt.
\newblock {An electron electric dipole moment search in the X 3Δ1 ground state
  of tungsten carbide molecules}.
\newblock {\em Journal of Modern Optics}, 56.

\bibitem{PhysRevA.52.3521}
J.~P. Jacobs, W.~M. Klipstein, S.~K. Lamoreaux, B.~R. Heckel, and E.~N.
  Fortson.
\newblock {Limit on the electric-dipole moment of $^{199}\mathrm{Hg}$ using
  synchronous optical pumping}.
\newblock {\em Phys. Rev. A}, 52:3521, 1995.

\bibitem{demille2001search}
D.~DeMille, F.~Bay, S.~Bickman, D.~Kawall, L.~Hunter, Krause D., S.~Maxwell,
  and K.~Ulmer.
\newblock {Search for the electric dipole moment of the electron using
  metastable PbO}.
\newblock In {\em AIP Conference Proceedings}, volume 596, pages 72--83, 2001.

\bibitem{PhysRevA.87.052130}
S.~Eckel, P.~Hamilton, E.~Kirilov, H.~W. Smith, and D.~DeMille.
\newblock {Search for the electron electric dipole moment using
  $\ensuremath{\Omega}$-doublet levels in PbO}.
\newblock {\em Phys. Rev. A}, 87:052130, 2013.

\end{thebibliography}

\end{document}